\documentclass[a4paper,manyauthors,nocleardouble,COMPASS]{cernphprep}

\usepackage{graphicx,hyperref,color,subfigure,xspace}
\usepackage{amssymb,amsmath}
\usepackage{wasysym}
\usepackage{lineno}

\newcommand{\ie}{\emph{i.e.}\xspace}

\newcommand{\eg}{\emph{e.g.}\xspace}

\DeclareMathOperator{\im}{Im}
\DeclareMathOperator{\re}{Re}
\newcommand{\beq}{\begin{equation}}
\newcommand{\eeq}{\end{equation}}

\begin{document}
\begin{titlepage}
\PHnumber{2017--xxx}
\PHdate{6 July 2017}
\title{New analysis of $\eta\pi$ tensor resonances measured at the COMPASS experiment}
\begin{abstract}
We present a new amplitude analysis of the $\eta\pi$ $D$-wave in $\pi^- p\to \eta\pi^- p$ measured by COMPASS. Employing an analytical model based on the principles of the relativistic $S$-matrix, we find two resonances that can be identified with the $a_2(1320)$ and the excited $a_2^\prime(1700)$, and perform a comprehensive analysis of their pole positions. For the mass and width of the  $a_2$ we find
   $M=(1307 \pm 1 \pm 6)$~MeV  and $\Gamma=(112 \pm 1 \pm 8)$~MeV, and for the excited state $a_2^\prime$
 we obtain $M=(1720 \pm 10 \pm 60)$~MeV and $\Gamma=(280\pm 10 \pm 70)$~MeV,  respectively.
\end{abstract}
 
\vspace*{60pt}
PACS: 14.40.Be; 11.55.Bq; 11.55.Fv; 11.80.Et; JLAB-THY-17-2468
\vfill
\Submitted{(to be submitted to Phys.\ Rev. Lett.)}
\newpage
%
%
\section*{JPAC Collaboration}
\label{app:jcollab}
\renewcommand\labelenumi{\textsuperscript{\theenumi}~}
\renewcommand\theenumi{\arabic{enumi}}
\begin{flushleft}
A.~Jackura\Irefnn{iu}{ceem}\Aref{jpa}\CorAuth, 
C.~Fern\'andez-Ram\'{\i}rez\Irefn{unam}\Aref{jpc},
M.~Mikhasenko\Irefn{bonniskp}\Aref{cm},
A.~Pilloni\Irefn{jlab}\Aref{jpa},
V.~Mathieu\Irefnn{iu}{ceem}\Aref{jpb},
J.~Nys\Irefn{ghent}\Aref{jpb}\Aref{jpd},
V.~Pauk\Irefn{jlab}\Aref{jpa},
A.~P.~Szczepaniak\Irefnnn{iu}{ceem}{jlab}\Aref{jpa}\Aref{jpb},
G.~Fox\Irefn{iucomp}\Aref{jpb},
\end{flushleft}

\section*{COMPASS Collaboration}
\label{app:collab}
\renewcommand\labelenumi{\textsuperscript{\theenumi}~}
\renewcommand\theenumi{\arabic{enumi}}
\begin{flushleft}
M.~Aghasyan\Irefn{triest_i},
R.~Akhunzyanov\Irefn{dubna}, 
M.G.~Alexeev\Irefn{turin_u},
G.D.~Alexeev\Irefn{dubna}, 
A.~Amoroso\Irefnn{turin_u}{turin_i},
V.~Andrieux\Irefnn{illinois}{saclay},
N.V.~Anfimov\Irefn{dubna}, 
V.~Anosov\Irefn{dubna}, 
A.~Antoshkin\Irefn{dubna}, 
K.~Augsten\Irefnn{dubna}{praguectu}
% ...
\end{flushleft}
%
%
\begin{Authlist}
\item \Idef{aveiro}{University of Aveiro, Dept.\ of Physics, 3810-193 Aveiro, Portugal}
\item \Idef{iu}{  Physics Dept., Indiana  University,  Bloomington,  IN  47405, USA}
\item \Idef{ceem}{Center for  Exploration  of  Energy  and  Matter, Indiana  University,  Bloomington,  IN  47403, USA}
\item \Idef{iucomp}{School of Informatics and Computing, Indiana University, Bloomington, IN 47405, USA}
\item \Idef{bochum}{Universit\"at Bochum, Institut f\"ur Experimentalphysik, 44780 Bochum, Germany\Arefs{l}\Aref{s}}
\item \Idef{bonniskp}{Universit\"at Bonn, Helmholtz-Institut f\"ur  Strahlen- und Kernphysik, 53115 Bonn, Germany\Arefs{l}}
\item \Idef{bonnpi}{Universit\"at Bonn, Physikalisches Institut, 53115 Bonn, Germany\Arefs{l}}
\item \Idef{brno}{Institute of Scientific Instruments, AS CR, 61264 Brno, Czech Republic\Arefs{m}}
\item \Idef{calcutta}{Matrivani Institute of Experimental Research \& Education, Calcutta-700 030, India\Arefs{n}}
\item \Idef{dubna}{Joint Institute for Nuclear Research, 141980 Dubna, Moscow region, Russia\Arefs{o}}
\item \Idef{ghent}{Dept.\ of Physics and Astronomy, Ghent University, 9000 Ghent, Belgium}
\item \Idef{freiburg}{Universit\"at Freiburg, Physikalisches Institut, 79104 Freiburg, Germany\Arefs{l}\Aref{s}}
\item \Idef{cern}{CERN, 1211 Geneva 23, Switzerland}
\item \Idef{liberec}{Technical University in Liberec, 46117 Liberec, Czech Republic\Arefs{m}}
\item \Idef{lisbon}{LIP, 1000-149 Lisbon, Portugal\Arefs{p}}
\item \Idef{mainz}{Universit\"at Mainz, Institut f\"ur Kernphysik, 55099 Mainz, Germany\Arefs{l}}
\item \Idef{unam}{Instituto de Ciencias Nucleares, Universidad Nacional Aut\'onoma de M\'exico,
                  Ciudad de M\'exico 04510, Mexico}
\item \Idef{miyazaki}{University of Miyazaki, Miyazaki 889-2192, Japan\Arefs{q}}
\item \Idef{moscowlpi}{Lebedev Physical Institute, 119991 Moscow, Russia}
\item \Idef{munichtu}{Technische Universit\"at M\"unchen, Physik Dept., 85748 Garching, Germany\Arefs{l}\Aref{r}}
\item \Idef{jlab}{Theory Center, Thomas  Jefferson  National  Accelerator  Facility, Newport  News,  VA  23606,  USA}
\item \Idef{nagoya}{Nagoya University, 464 Nagoya, Japan\Arefs{q}}
\item \Idef{praguecu}{Charles University in Prague, Faculty of Mathematics and Physics, 18000 Prague, Czech Republic\Arefs{m}}
\item \Idef{praguectu}{Czech Technical University in Prague, 16636 Prague, Czech Republic\Arefs{m}}
\item \Idef{protvino}{State Scientific Center Institute for High Energy Physics of National Research Center `Kurchatov Institute', 142281 Protvino, Russia}
\item \Idef{saclay}{IRFU, CEA, Universit\'e Paris-Saclay, 91191 Gif-sur-Yvette, France\Arefs{s}}
\item \Idef{taipei}{Academia Sinica, Institute of Physics, Taipei 11529, Taiwan\Arefs{tw}}
\item \Idef{telaviv}{Tel Aviv University, School of Physics and Astronomy, 69978 Tel Aviv, Israel\Arefs{t}}
\item \Idef{triest_u}{University of Trieste, Dept.\ of Physics, 34127 Trieste, Italy}
\item \Idef{triest_i}{Trieste Section of INFN, 34127 Trieste, Italy}
\item \Idef{turin_u}{University of Turin, Dept.\ of Physics, 10125 Turin, Italy}
\item \Idef{turin_i}{Torino Section of INFN, 10125 Turin, Italy}
\item \Idef{tomsk}{Tomsk Polytechnic University,634050 Tomsk, Russia\Arefs{nauka}}
\item \Idef{illinois}{University of Illinois at Urbana-Champaign, Dept.\ of Physics, Urbana, IL 61801-3080, USA\Arefs{nsf}}
\item \Idef{warsaw}{National Centre for Nuclear Research, 00-681 Warsaw, Poland\Arefs{u}}
\item \Idef{warsawu}{University of Warsaw, Faculty of Physics, 02-093 Warsaw, Poland\Arefs{u}}
\item \Idef{warsawtu}{Warsaw University of Technology, Institute of Radioelectronics, 00-665 Warsaw, Poland\Arefs{u} }
\item \Idef{yamagata}{Yamagata University, Yamagata 992-8510, Japan\Arefs{q} }
\end{Authlist}
%
%
\renewcommand\theenumi{\alph{enumi}}
\begin{Authlist}
\item [{\makebox[2mm][l]{\textsuperscript{\#}}}] Corresponding authors
\item [{\makebox[2mm][l]{\textsuperscript{*}}}] Deceased
\item \Adef{jpa}{Supported by U.S.~Dept.\ of Energy, Office of Science,
                 Office of Nuclear Physics under contracts DE-AC05-06OR23177, DE-FG0287ER40365}
\item \Adef{jpc}{Supported by PAPIIT-DGAPA (UNAM, Mexico) Grant No.~IA101717, by CONACYT (Mexico)
                 Grant No.~251817 and by Red Tem\'atica CONACYT de F\'{\i}sica en Altas Energ\'{\i}as
                 (Red FAE, Mexico)}
\item \Adef{cm}{ Also a member of the COMPASS Collaboration}
\item \Adef{jpb}{Supported by National Science Foundation Grant PHY-1415459}
\item \Adef{jpd}{Supported as an `FWO-aspirant' by the Research Foundation Flanders (FWO-Flanders)}
\item \Adef{a}{Also at Instituto Superior T\'ecnico, Universidade de Lisboa, Lisbon, Portugal}
\item \Adef{b}{Also at Dept.\ of Physics, Pusan National University, Busan 609-735, Republic of Korea and at Physics Dept., Brookhaven National Laboratory, Upton, NY 11973, USA}
\item \Adef{ictp}{Also at Abdus Salam ICTP, 34151 Trieste, Italy}
\item \Adef{r}{Supported by the DFG cluster of excellence `Origin and Structure of the Universe' (www.universe-cluster.de) (Germany)}
\item \Adef{p2i}{Supported by the Laboratoire d'excellence P2IO (France)}
\item \Adef{d}{Also at Chubu University, Kasugai, Aichi 487-8501, Japan\Arefs{q}}
\item \Adef{x}{Also at Dept.\ of Physics, National Central University, 300 Jhongda Road, Jhongli 32001, Taiwan}
\item \Adef{lpr}{Present address: LP-Research Inc., Tokyo, Japan}
\item \Adef{e}{Also at KEK, 1-1 Oho, Tsukuba, Ibaraki 305-0801, Japan}
\item \Adef{g}{Also at Moscow Institute of Physics and Technology, Moscow Region, 141700, Russia}
\item \Adef{h}{Present address: RWTH Aachen University, III.\ Physikalisches Institut, 52056 Aachen, Germany}
\item \Adef{o}{    Supported by CERN-RFBR Grant 12-02-91500}
\item \Adef{lpr}{Present address: LP-Research Inc., Tokyo, Japan}
\item \Adef{y}{Also at Dept.\ of Physics, National Kaohsiung Normal University, Kaohsiung County 824, Taiwan}
\item \Adef{turin_p}{Also at University of Eastern Piedmont, 15100 Alessandria, Italy}
\item \Adef{i}{Present address: Uppsala University, Box 516, 75120 Uppsala, Sweden}
\item \Adef{c}{    Supported by the DFG Research Training Group Programmes 1102 and 2044 (Germany)} 
%
%
\item \Adef{l}{    Supported by BMBF - Bundesministerium f\"ur Bildung und Forschung (Germany)}
\item \Adef{s}{    Supported by FP7, HadronPhysics3, Grant 283286 (European Union)}
\item \Adef{m}{    Supported by MEYS, Grant LG13031 (Czech Republic)}
\item \Adef{n}{    Supported by SAIL (CSR) and B.Sen fund (India)}
\item \Adef{p}{\raggedright 
                   Supported by FCT - Funda\c{c}\~{a}o para a Ci\^{e}ncia e Tecnologia, COMPETE and QREN, Grants CERN/FP 116376/2010, 123600/2011 
                   and CERN/FIS-NUC/0017/2015 (Portugal)}
\item \Adef{q}{    Supported by MEXT and JSPS, Grants 18002006, 20540299, 18540281 and 26247032, the Daiko and Yamada Foundations (Japan)}
\item \Adef{tw}{   Supported by the Ministry of Science and Technology (Taiwan)}
\item \Adef{t}{    Supported by the Israel Academy of Sciences and Humanities (Israel)}
\item \Adef{nauka}{Supported by the Russian Federation  program ``Nauka'' (Contract No. 0.1764.GZB.2017) (Russia)}
\item \Adef{nsf}{  Supported by the National Science Foundation, Grant no. PHY-1506416 (USA)}
\item \Adef{u}{    Supported by NCN, Grant 2015/18/M/ST2/00550 (Poland)}
\end{Authlist}

\end{titlepage}

\section{Introduction}
The spectrum of hadrons contains a number of poorly 
determined or missing resonances, 
the better knowledge of which is of key importance for improving our understanding of 
Quantum Chromodynamics (QCD).
Active research programs in this 
direction are being pursued at various experimental facilities, 
including the COMPASS and LHCb experiments at 
CERN~\cite{Abbon:2014aex,Adolph:2014rpp,Adolph:2015tqa,Alves:2008zz}, 
CLAS/CLAS12 and GlueX at JLab~\cite{Glazier:2015cpa,Ghoul:2015ifw,AlGhoul:2017nbp}, 
BESIII at BEPCII~\cite{Fang:2016gsh}, BaBar, and Belle~\cite{Bevan:2014iga}. 
To connect the experimental observables 
with the QCD predictions, an amplitude analysis is required. 
Fundamental principles of $S$-matrix theory, such as unitarity 
and analyticity (which originate from probability conservation and causality), 
should be applied in order to construct reliable reaction models. 
When resonances dominate the spectrum, which is the case studied here, unitarity is especially  important since
 it constrains resonance widths and allows us to determine the location of resonance poles in the 
complex energy plane of the multivalued partial wave amplitudes. 

In 2014, COMPASS published high-statistics partial wave analyses of the 
$\pi^- p\to \eta^{(\prime)}\pi^- p$ reaction, at 
$p_\text{beam} = 191$~GeV~\cite{Adolph:2014rpp}. 
The odd angular-momentum waves have exotic quantum numbers and 
exhibit structures that may be compatible with a hybrid meson~\cite{Meyer:2015eta}. The even waves show strong signals of non-exotic resonances. In particular, the $D$-wave of $\eta\pi$, with $I^G(J^{PC}) = 1^-(2^{++})$, is dominated by the peak of the $a_2(1320)$ and its 
Breit-Wigner parameters were extracted and presented in Ref.~\cite{Adolph:2014rpp}. The $D$-wave also exhibits a hint of the first radial excitation, the $a_2^\prime(1700)$~\cite{pdg}. 

In this letter we present a new analysis of the $D$-wave  based on an analytical model constrained by unitarity, which extends beyond a simple  Breit-Wigner parameterization. Our model builds on a more general framework for  a systematic analysis of peripheral meson production, which is currently under development \cite{Mikhasenko:2017jtg,Jackura:2016llm,Nys:2016vjz}. 
Using the 2014 COMPASS measurement as input, the model is fitted to the results of the mass-independent analysis that was performed in 40 MeV wide bins of the $\eta\pi$ mass. The $a_2$ and $a_2^{\prime}$ resonance parameters are extracted in the single-channel approximation and the coupled-channel effects are estimated by including the $\rho\pi$ final state. We determine the statistical uncertainties by means of the bootstrap
method~\cite{recipes,Fernandez-Ramirez:2015tfa,Blin:2016dlf,Landay:2016cjw,zc3900}, 
and assess the systematic uncertainties in the pole positions 
by varying model-dependent parameters in the reaction amplitude.

To the best of our knowledge, this is the first precision determination 
of pole parameters of these resonances that includes the recent, most precise, COMPASS data. 

\section{Reaction Model}
\begin{figure}[t]
\centering
\subfigure[Reaction diagram.]{
\includegraphics[width=.48\columnwidth]{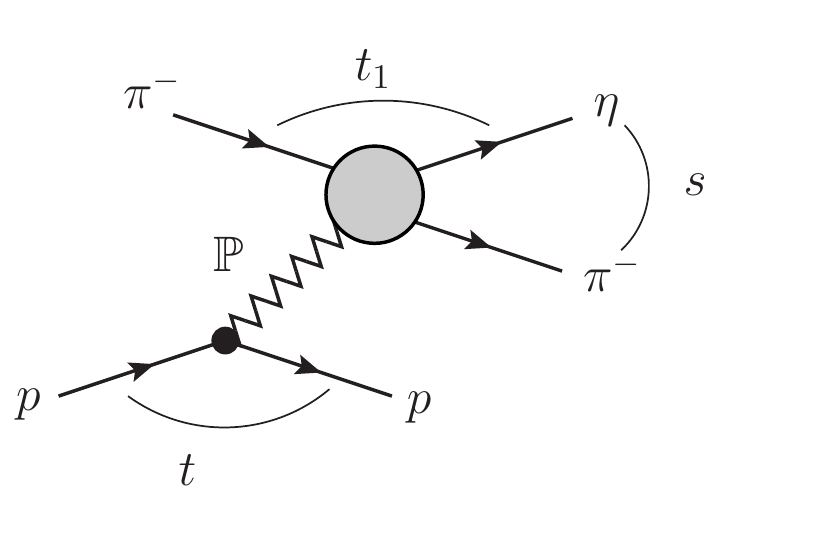}\label{fig:rxn}}
\subfigure[Unitarity diagram.]{
\includegraphics[width=.78\columnwidth]{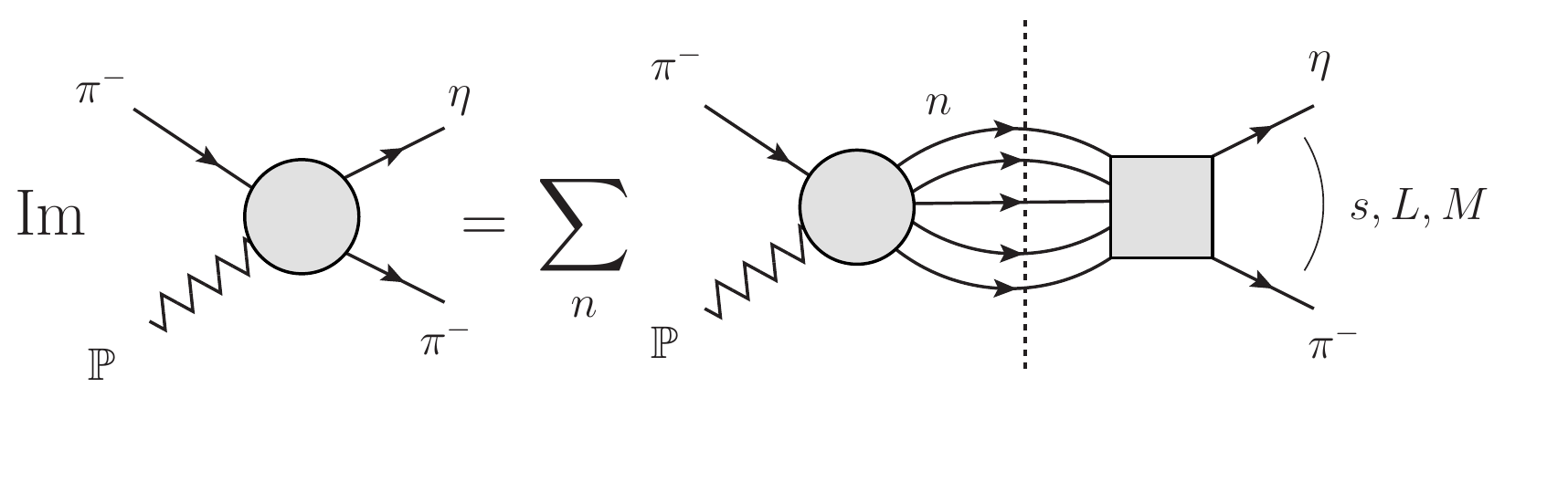}\label{fig:etapi_unit}
}
\caption{(a) Pomeron exchange in $\pi^- p \to \eta\pi^- p$.
(b) 
The $\pi\mathbb{P}\to \eta\pi$ amplitude is expanded in partial waves in the $s$-channel of the $\eta\pi$ system, $a_{JM}(s)$, with $J = L$ and $t \to t_{\text{eff}}$. 
Unitarity relates the imaginary part of the amplitude to final state interactions that include all kinematically allowed intermediate states $n$.
}
\label{fig:etapi_diag}  
\end{figure}

We consider the peripheral production process $\pi p\to \eta\pi p$ (Fig.~\ref{fig:rxn}), which is dominated by Pomeron ($\mathbb{P}$) exchange. High-energy diffractive production allows us to assume factorization of the ``top" vertex, so that the $\pi\mathbb{P}\to \eta\pi$ amplitude resembles an  ordinary helicity amplitude~\cite{Hoyer:1974hd}. It is a function of $s$ and $t_1$, 
 the $\eta\pi$ invariant mass squared and the invariant momentum  transfer squared between the incoming pion and the $\eta$, respectively. It also depends on $t$, the momentum transfer between the nucleon target and recoil. In the Gottfried-Jackson (GJ) frame~\cite{Gottfried:1964nx}, the Pomeron helicity in $\pi\mathbb{P}\to \eta\pi$  equals the $\eta\pi$ total angular momentum projection $M$,  
and the helicity amplitudes $a_M(s,t,t_1)$ can be expanded in partial waves $a_{JM}(s,t)$  with total angular momentum $J = L$. 
   The allowed quantum numbers of the $\eta\pi$ partial waves  are $J^{P}=1^-$, $2^+$, $3^-,\ldots$. The Pomeron exchange has natural parity. Parity relates the amplitudes with opposite spin projections $a_{JM} = -a_{J-M}$~\cite{Chung:1974fq}. That is, the $M=0$ amplitude is forbidden and the two $M=\pm 1$ amplitudes are given, up to a sign, by a single scalar function.

The assumption about the Pomeron dominance can be  
quantified by the magnitude of unnatural partial waves. 
In the analysis of ref.~\cite{Adolph:2014rpp}, the magnitude of the $L=M=0$ wave was estimated to 
be $<1\%$, and it also absorbs other possible reducible backgrounds. 
The patterns of azimuthal dependence in the central production of mesons~\cite{Kirk:1998et, Barberis:1999zh, Barberis:1998bq, Barberis:1998ax,Close:1997pj} indicate that at low momentum transfer, $t\sim 0$, the Pomeron behaves as a vector~\cite{Close:1999is, Arens:1996xw}, which is  
 in  agreement with the strong dominance of the $|M|=1$ component in the COMPASS data.~\footnote{At low $t$, the Pomeron trajectory passes through $J=1$, while at larger, positive $t$, the trajectory is expected to pass though $J=2$ where it would  relate to the tensor glueball~\cite{Lebiedowicz:2013ika,Lebiedowicz:2016ioh}.}
We are unable to further address the nature of the exchange from the 
data of ref.~\cite{Adolph:2014rpp}
since they are 
integrated over the momentum transfer $t$~\footnote{For example, Ref.~\cite{Bromberg:1983cu} suggested a dominance of $f_2$ exchanges for $a_2(1320)$ production. To probe this, one should analyze the $t$ and total energy dependences.}. However, from this single energy analysis we cannot be sure the exchange is purely Pomeron. Analyses such as Ref. \cite{Bromberg:1983cu} suggest there could be an $f$ exchange, but for our analysis the natural parity exchanges will be similar, so we consider an effective Pomeron which may be a mixture of pure Pomeron and $f$.
We note here that COMPASS has published data in the $3\pi$ channel, which are binned both in $3\pi$ invariant mass and momentum transfer $t$ \cite{Adolph:2015tqa}.

The COMPASS mass-independent analysis~\cite{Adolph:2014rpp} is restricted to partial waves with $L=1$ to $6$ and $|M|=1$ 
 (except for the $L=2$ where also the $|M|=2$ wave is taken into account). 
The lowest mass exchanges in the crossed channels of 
$\pi\mathbb{P}\to \eta\pi$ correspond to the $a$ (in the $t_1$ channel) and the $f$ (in the $u_1$ channel) trajectories, 
thus higher partial waves are not expected to be significant 
in the $\eta\pi$ mass region of interest, $\sqrt{s} < 2$~GeV. 
However, the systematic uncertainties associated with an analysis based 
on a truncated set of partial waves is hard to estimate. 
 
To compare with the partial wave intensities measured in Ref.~\cite{Adolph:2014rpp}, which are integrated over $t$ from $t_{\text{min}}=-1.0$~GeV$^2$ to $t_{\text{max}}=-0.1$~GeV$^2$, we use an effective value for the momentum transfer 
$t_{\text{eff}}=-0.1$~GeV$^2$ and $a_{JM}(s) \equiv a_{JM}(s,t_{\text{eff}})$. The effect of a possible $t_{\text{eff}}$ dependence is taken into account in the estimate of the systematic uncertainties. The natural parity exchange partial wave amplitudes $a_{JM}(s)$ can be identified  with the  amplitudes $A^{\epsilon=1}_{LM}(s)$ as defined in Eq.~(1) of Ref.~\cite{Adolph:2014rpp}, where $\epsilon=+1$ is the reflectivity eigenvalue that selects the natural parity exchange.

In the following we consider the single,  
$J=2$, $|M|=1$ natural parity partial wave, which
we denote by $a(s)$, and fit its modulus squared to the measured 
(acceptance corrected) number of events~\cite{Adolph:2014rpp}:
\beq\label{eq:Intensity}
\frac{d\sigma}{d \sqrt{s}} \propto I(s)  
=  \int_{t_\text{min}}^{t_\text{max}} \,dt\, p\,  \left| a(s,t) \right|^2 
\equiv  \mathcal{N} p\,\left| a(s) \right|^2.
\eeq
Here, $I(s)$ is the intensity distribution of the $D$ wave,  $p=\lambda^{1/2}(s,m_{\eta}^2,m_{\pi}^2)/(2\sqrt{s})$ 
the $\eta\pi$ breakup momentum, and  
 $q=\lambda^{1/2}(s,m_{\pi}^2,t_{\text{eff}})/(2\sqrt{s})$, which will be used later,  is the $\pi$ beam momentum in the $\eta\pi$ rest frame with 
$\lambda(x,y,z) =x^2 + y^2 + z^2 -2xy -2xz -2 yz$ being the K\"all\'en triangle function.
Since the physical normalization of the cross section is not determined in Ref.~\cite{Adolph:2014rpp},  
the constant $\mathcal{N}$ on the right hand side of Eq.~\eqref{eq:Intensity} is a free parameter.

In principle, one should consider the coupled-channel problem involving all the kinematically allowed intermediate states (see Fig.~\ref{fig:etapi_unit}). 
The PDG reports the important final states for the $2^{++}$ system are the $3\pi$ ($\rho\pi$, $f_2\pi$) and $\eta\pi$ systems~\cite{pdg}.
Far from thresholds, a narrow peak in the data is generated by a pole in the closest unphysical sheet, 
regardless of the number of open channels. 
The residues (related to the branching ratios) depend on the individual couplings of each channel to the resonance, 
and therefore their extraction requires the inclusion of all the relevant channels.
However, the pole position is expected to be essentially insensitive to the inclusion of multiple channels. This is easily understood in the Breit-Wigner approximation, where the total width 
extracted for a given state is 
independent of the branchings to individual channels.  
Thus, when investigating the pole position, we restrict the analysis to the elastic approximation, where only $\eta\pi$ can appear in the intermediate state. We will elaborate on the effects of introducing the $\rho\pi$ channel, which is known to be the dominant one of the decay of $a_2(1320)$~\cite{pdg}, as part of the systematic checks.

In the resonance region, unitarity gives constraints for both the $\eta\pi$ 
interaction and production. Denoting the $\eta\pi \to \eta\pi$ 
scattering $D$-wave by $f(s)$, 
unitarity and analyticity determine the imaginary part of both 
amplitudes above the $\eta\pi$ threshold $s_{th} = (m_\eta + m_\pi)^2$: 
\begin{align}
\im\hat a(s)  & = \rho(s)\,\hat f^{\,*}(s)\,\hat a(s) \label{eq:InelasUnit}, \\
\im\hat f(s) &= \rho(s)\,\lvert \hat f(s) \rvert^2 \label{eq:ElasUnit}.
\end{align}
From  the analysis of kinematical singularities~\cite{Shimada:1978sx,Danilkin:2014cra,Gribov:2009zz}
it follows that the amplitude $a(s)$ appearing in Eq.~\eqref{eq:Intensity} has kinematical singularities proportional to $K(s)=p^2q$, and $f(s)$ has singularities proportional to $p^4$ . 
The reduced partial waves in Eqs.\eqref{eq:InelasUnit} and \eqref{eq:ElasUnit} are free from kinematical singularities, and defined by  
\eg $\hat a(s) = a(s)/K(s)$, $\hat f(s) = f(s)/p^{4}$, with  $\rho(s)=2p^{5}/\!\sqrt{s}$ 
being the two-body phase space factor 
that absorbs the barrier factors of the $D$-wave. Note that Eq.~\eqref{eq:InelasUnit} is the elastic approximation of Fig.~\ref{fig:etapi_unit}.

We write $\hat{f}$ in the standard N-over-D form,  $\hat{f}(s)=N(s)/D(s)$, with $N(s)$ absorbing singularities from exchange interactions, \ie ``forces'' acting between $\eta\pi$  also known as left hand cuts,  
and $D(s)$ containing the right hand cuts that are associated with direct channel thresholds. 
Unitarity leads to a relation between $D$ and $N$, $\im D(s) = -\rho(s)N(s)$, with the general once-subtracted integral solution
\beq\label{eq:Dsol}
D(s) =  D_{0}(s) - \frac{s}{\pi}\int_{s_{th}}^{\infty}ds'\frac{\rho(s')N(s')}{s'(s'-s)}. 
\eeq
Here, the function $D_0(s)$ 
is real for $s>s_{th}$ and can be parameterized as
\beq\label{eq:DCDD}
D_0(s) = c_0-c_1s - \frac{c_2}{c_3-s}.
\eeq
Note that the subtraction constant has been absorbed into $c_0$ of $D_0(s)$. The rational function in Eq.~\eqref{eq:DCDD} is a sum over two so-called Castillejo-Dalitz-Dyson (CDD) poles \cite{Castillejo:1955ed}, with the first pole located at $s = \infty$   (CDD$_{\infty}$) and the second one at $s = c_3$. The CDD poles produce real zeros of the amplitude $\hat f$ and they also lead to poles of $\hat f$ on the complex plane (second sheet). Since these poles are introduced via parameters  like $c_1$, $c_2$, rather than being generated through $N$ ({\it cf.}  Eq.~\eqref{eq:Dsol}),  they  are commonly attributed to genuine QCD states, \ie states that  do not originate from effective, long-range interactions  such as pion exchange ~\cite{frautschi1963regge}. 
In order to fix the arbitrary normalization of $N(s)$ and $D(s)$, 
we set $c_0=(1.23)^2$ since it is expected to 
be numerically close to the $a_2$ mass squared expressed in GeV. 
One also expects $c_1$ to be approximately 
equal to the slope of the leading Regge 
trajectory~\cite{Londergan:2013dza}. 
The quark model~\cite{Godfrey:1985xj} and lattice QCD~\cite{Dudek:2011tt} 
predict two states in the energy region of interest, so we use only two CDD poles. 
It follows from Eq.~\eqref{eq:Dsol} that the singularities of $N(s)$ 
(which originate from the finite range of the interaction) 
 will also appear on the second sheet in $D(s)$, 
together with the resonance poles generated by the CDD terms.
We use a simple model for $N(s)$, where the left hand cut is approximated 
by a higher order pole, 
\beq
\rho(s) N(s)=g \,\frac{\lambda^{5/2}(s,m_\eta^2,m_\pi^2)}{(s+s_R)^{n}}.
\eeq
Here, $g$ and $s_R$ effectively parameterize 
the strength and inverse range of the exchange 
forces in the $D$-wave, whereas the power $n=7$ makes the integral 
in Eq.~\eqref{eq:Dsol} account for the finite range of interactions with the appropriate powers to regulate the threshold singularities, and additional powers as the model for the left hand singularities.
The parameterization of $N(s)$ removes the kinematical $1/s$ singularity in $\rho(s)$. 
Therefore, dynamical singularities on the second sheet are
either associated with the particles represented by the CDD poles, 
or the exchange forces parameterized by the higher order pole in $N(s)$.

 The general parameterization 
for $\hat{a}(s)$, which is constrained by unitarity in Eq.~\eqref{eq:InelasUnit},  is obtained following similar arguments and is given by a ratio of two functions
\begin{equation} \label{eq:nD}
\hat{a}(s) = \frac{n(s)}{D(s)},
\end{equation} 
where $D(s)$ is given by Eq.~\eqref{eq:Dsol} and  
 brings in the effects of $\eta\pi$ final state interactions, while $n(s)$ describes the exchange interactions in the production process $\pi \mathbb{P} \to \eta\pi$ and contains the associated left hand singularities.  In both the production process and the  
elastic scattering no important contributions from light meson exchanges are expected since the lightest resonances in the $t_1$  and $u_1$ channels are the $a_2$ and $f_2$ mesons, respectively. Therefore, the numerator function in   Eq.~\eqref{eq:nD} is expected to be a smooth function of $s$ in the complex plane near the physical region, with one exception: 
the CDD pole at $s=c_3$ produces a zero in $\hat{a}(s)$. Since a zero in the elastic scattering amplitude does not in general imply a zero in the production amplitude,  we write $n(s)$ as
\beq\label{eq:neqn}
n(s) = \frac{1}{c_3-s}\sum_{j}^{n_p}\,a_{j}\,T_{j}(\omega(s)),
\eeq
where the function to the right of the pole is expected to be 
 analytical in $s$ near the physical region.  
We parameterize it using the Chebyshev polynomials $T_j$, with 
$\omega(s)=s/(s+\Lambda)$ approximating the left hand singularities 
in the production process, 
$\pi\mathbb{P}\to \eta\pi$. The real coefficients 
$a_j$ are determined from the fit to the data. 
In the analysis, we fix $\Lambda=1$ GeV$^2$. 
We choose an expansion in Chebyshev polynomials 
as opposed to a simple power series in $\omega$ 
to reduce the correlations between the $a_j$ parameters. Since we examine the partial wave intensities integrated over the momentum transfer  $t$, we assume that the expansion coefficients are independent of $t$. The only $t$-dependence comes from the residual  kinematical dependence on the breakup momentum $q$.
 
A comment on the relation between the N-over-D method and the  $K$-matrix parameterization is worth noting.  
If one assumes that there are no left hand singularities, \ie let $N(s)$ be a constant,
then Eq.~\eqref{eq:Dsol} is identical to that of the standard 
$K$-matrix formalism~\cite{Aitchison:1972ay}.
Hence we can relate both approaches through $K^{-1}(s) = D_0(s)$. 
It is also worth noting that  the parameterization in  Eq.~\eqref{eq:DCDD} automatically
satisfies  causality, \ie there are no poles on the physical energy sheet. 

\begin{figure}
\centering
\subfigure[CDD$_{\infty}$ pole only.]{
\includegraphics[width=0.45\columnwidth]{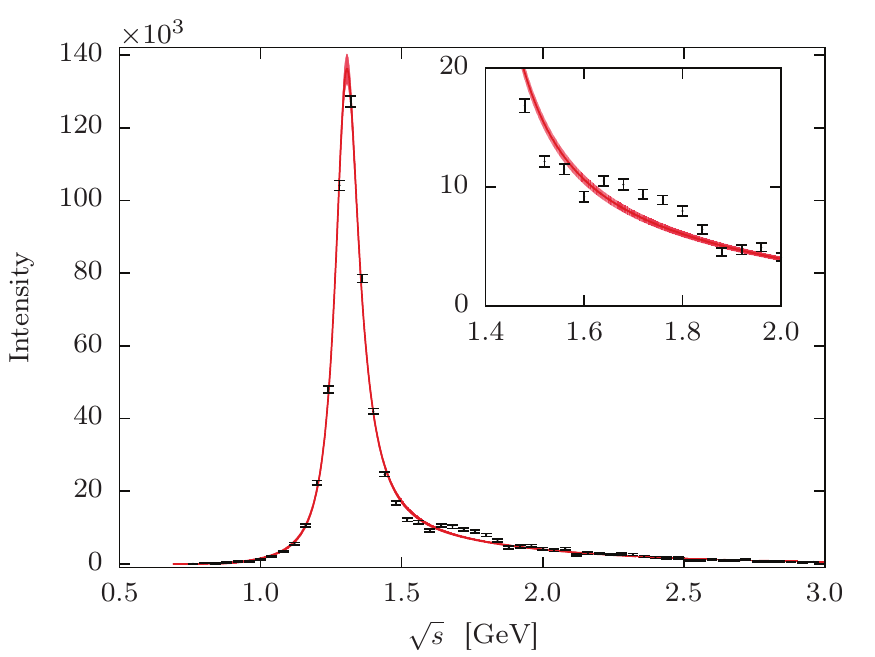}
\label{fig:1CDD}}
\subfigure[Two CDD poles.]{
\includegraphics[width=0.45\columnwidth]{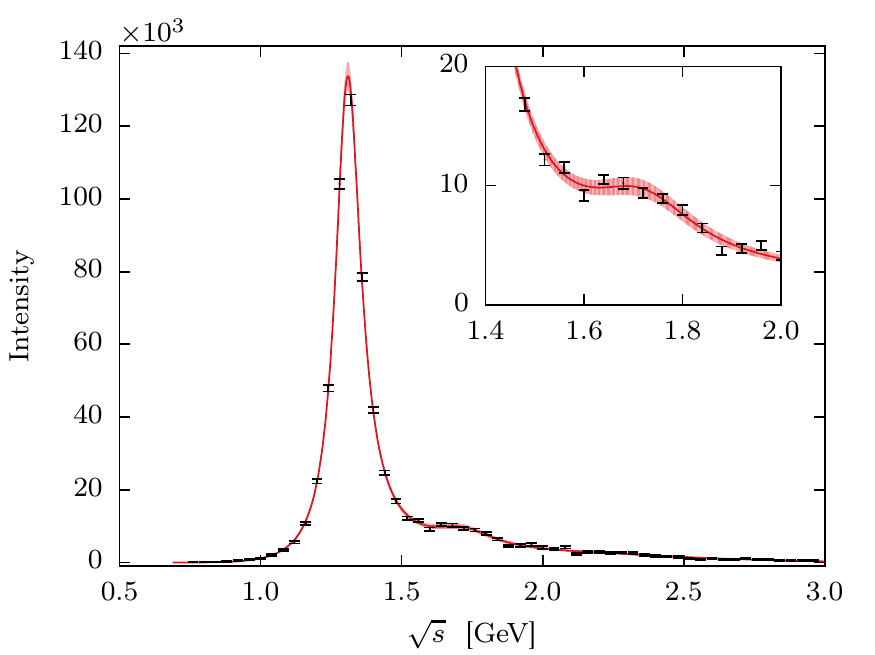}
\label{fig:2CDD}}
\caption{Intensity distribution and fits to the $J^{PC}=2^{++}$ wave  
for different number of CDD poles, (a) using only CDD$_{\infty}$  and (b) 
 using CDD$_{\infty}$ and the CDD pole at $s=c_3$. Red lines are fit results with  
$I(s)$ given by Eq.~\eqref{eq:Intensity}. Data is taken from Ref.~\cite{Adolph:2014rpp}. The inset shows the $a'_2$ region. 
The error bands correspond to the $3\sigma$ ($99.7\%$) confidence level.}
\label{fig:intensities}
\end{figure}
%

\begin{table}[tb]
\centering
\caption{Best fit denominator and production parameters for the fit with two CDD poles, $s_R=1.5$ GeV$^2$, 
$\mathcal{N}=10^6$, $c_0=(1.23)^2$, and the number of expansion 
parameters $n_p=6$, leading to $\chi^2/\text{d.o.f.}=1.91$.  
 Denominator uncertainties are determined from a bootstrap analysis using $10^5$ random fits. We report no uncertainties on the production parameters as they are highly correlated.} \label{tab:params}
\begin{tabular}{ccc|cc}
\multicolumn{3}{c|}{Denominator parameters} 
&\multicolumn{2}{c}{Production parameters\,\, [GeV$^{-2}$]}  \\
 \hline
$c_1$ & $\phantom{00}0.532 \pm 0.006$\,\,&\, GeV$^{-2}$ & $a_0$ 
&  $\phantom{-}0.471$  \\
$c_2$ & $\phantom{00}0.253 \pm 0.007$\,\,& GeV$^{2}$  & $a_1$ & $\phantom{-}0.134 $ \\
$c_3$ & $\phantom{00}2.38 \pm 0.02$\,\,& GeV$^{2}$ & $a_2$ &   $-1.484 $ \\
$g$ & $113 \pm 1$\,\,& GeV$^{4}$ & $a_3$ & $\phantom{-}0.879$ \\
 &  &  & $a_4$ & $\phantom{-}2.616$ \\
 &  &  & $a_5$ & $-3.652$  \\
 &  &  & $a_6$ & $\phantom{-}1.821$\\
\end{tabular}
\end{table}
\begin{figure}
\centering
\includegraphics[width=.5\textwidth]{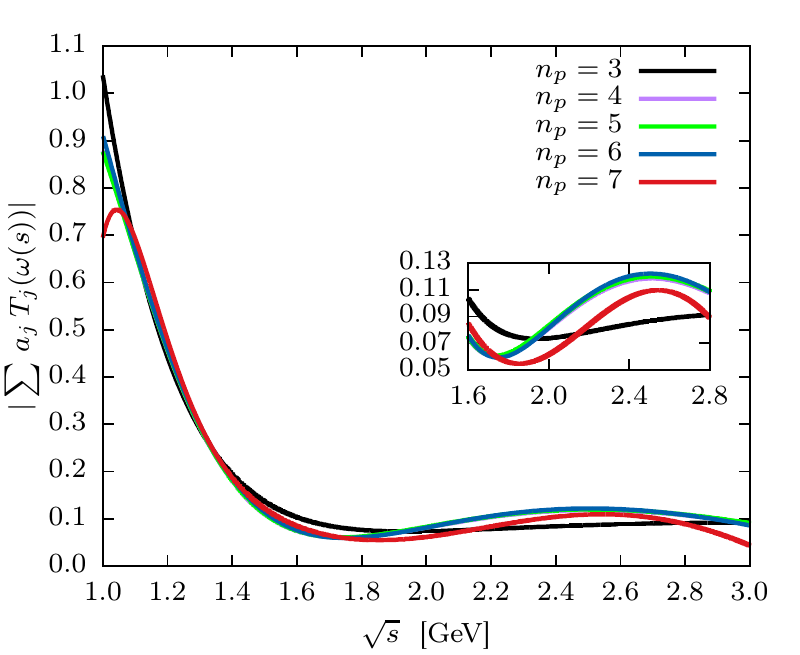}
\caption{Amplitude numerator function $\lvert\sum_j^{n_p}\,a_j\,T_j(\omega(s))\rvert$ 
for different values of $n_p$. The absolute value is taken as there is a phase ambiguity because we fit only the intensity $\sim \lvert a(s)\rvert^2$. Note that each curve is an independent fit for a specific number of terms $n_p$. The curves for $n_p=4$, 5, and 6 all coincide in the resonance region, as shown in the inset.
}
\label{fig:prod}  
\end{figure}
%

\section{Methodology}
We fit our model to the intensity distribution
for $\pi^- p\to \eta\pi^- p$ in the $D$-wave (56 data points)~\cite{Adolph:2014rpp}, 
as defined in Eq.~\eqref{eq:Intensity}, by minimizing $\chi^2$. 
We fix the overall scale, $\mathcal{N} = 10^6$ (see Eq.~\eqref{eq:Intensity}), and fit the coefficients $a_j$ (see Eq.~\eqref{eq:neqn}), which are then expected to be $O(1)$,  
 and also the parameters in the $D(s)$ function.  
In the first step we obtain the best fit for a given total number of parameters, 
and in the second step we estimate the statistical uncertainties  
 using the bootstrap 
technique~\cite{recipes,Fernandez-Ramirez:2015tfa,Blin:2016dlf,Landay:2016cjw,zc3900}. 
That is to say, we generate $10^5$ pseudodata 
sets, each data point being resampled according to a Gaussian 
distribution having as mean and standard deviation the original value and error, 
and we repeat the fit for each set. 
In this way, we obtain $10^5$ different values for the fit parameters, 
and we take the means and standard deviations 
as expected values and statistical uncertainties, respectively. 
   
In order to assess the systematic uncertainties   
we study the dependence of the pole parameters on variations of the model, specifically we change
$i)$ the number of CDD poles from 1 to 3, 
$ii)$ the total number of terms $n_p$ in the expansion of the numerator function $n(s)$ in Eq.~\eqref{eq:neqn}, 
$iii)$ the value of $s_R$ in the left hand cut model, 
$iv)$ the value of $t_\text{eff}$ of the total momentum transfered,  
and $v)$ the addition of the $\rho\pi$ channel to study coupled-channel effects.  

The fit with CDD$_\infty$ only, 
shown in Fig.~\ref{fig:1CDD}, for $s_R=1.5$ GeV$^2$ and $n_p=6$ (with a total of 9 parameters), captures neither the dip at $1.5$~GeV nor the bump at $1.7$~GeV. 
In contrast, the fit with two CDD poles (11 parameters),  
shown in Fig.~\ref{fig:2CDD}, captures both features, 
giving a $\chi^2/\text{d.o.f.} = 86.17/(56-11) = 1.91$. 
The parameters corresponding to the best fit with two CDD poles are given 
in Table~\ref{tab:params}.
The addition of another CDD pole does not improve the fit, 
as the limited precision in the data is incapable of indicating any further resonances. Specifically  the residue of the additional pole turns out to be compatible with zero, 
 leaving the other fit parameters unchanged. 
We associate no systematic uncertainty to that.

As discussed earlier, an acceptable numerator function  $n(s)$ should be 
``smooth" in the resonance region, 
\ie  without significant peaks or dips on the scale of the resonance widths. 
The parameters $c_i$ and $g$ of the denominator function are related 
to resonance parameters, while $s_R$ controls the distant 
second sheet singularities due to exchange forces. 
The expansion in $n(s)$, shown in Fig.~\ref{fig:prod} 
for $s_R=1.5$ GeV$^2$ and two CDD poles, 
has a singularity occurring at $s=-1.0$ GeV$^2$ 
because of the definition of $\omega(s)$ and our choice of $\Lambda$ \footnote{Note that the production term is not well constrained below $s\sim 1$ GeV$^2$, as the phase space and barrier factors highly suppress the near threshold behavior. The singularity at $s=-1$ GeV$^2$ however, persist for each $n_p$ solution.}. For variations in $n(s)$ between $n_p=3$ and $n_p=7$, we find the pole positions are relatively stable, which we discuss later in our systematic estimates.

The dependence on $t_{\text{eff}}$ is expected to affect mostly the overall normalization. Indeed, the variation from $t_{\text{eff}}=-1.0$ GeV$^2$ to $-0.1$ GeV$^2$ 
gives less than $2\%$ difference for the $a_2^\prime(1700)$ parameters, and $<1\permil$ for the $a_2(1320)$, and can be neglected compared to the other uncertainties.

\section{Results}
\begin{figure}
\centering
\includegraphics[width=.5\textwidth]{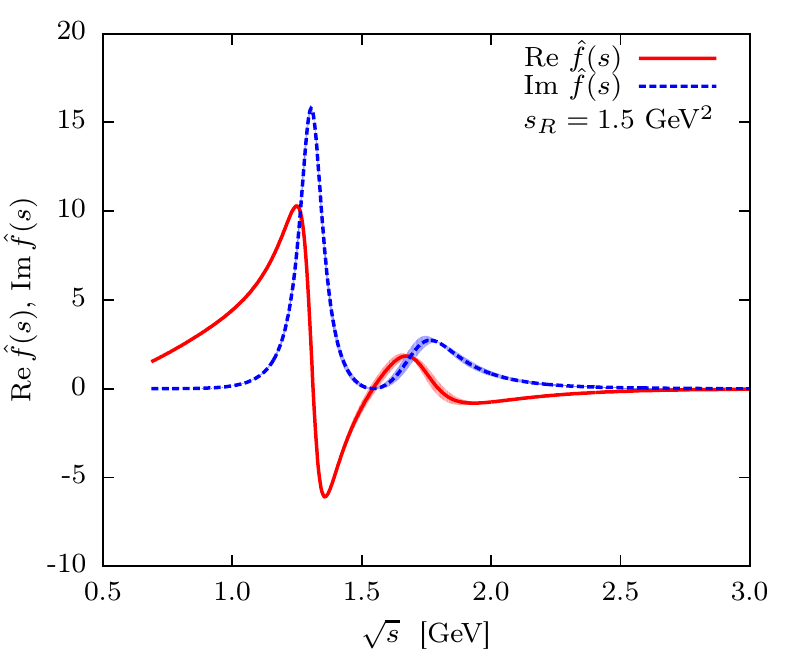}
\caption{The reduced $\eta\pi\to\eta\pi$ partial amplitude in the $D$-wave, $\hat{f}(s) = N(s)/D(s)$. 
Shown are the real (red) and imaginary (blue) 
parts as a function of the $\eta\pi$ invariant mass with $3\sigma$ error band. 
The node in the imaginary part at $1.7$~GeV 
is due to the total correlation between the real and imaginary parts. 
}
\label{fig:etapi}  
\end{figure}
\begin{figure*}
\includegraphics[width=\textwidth]{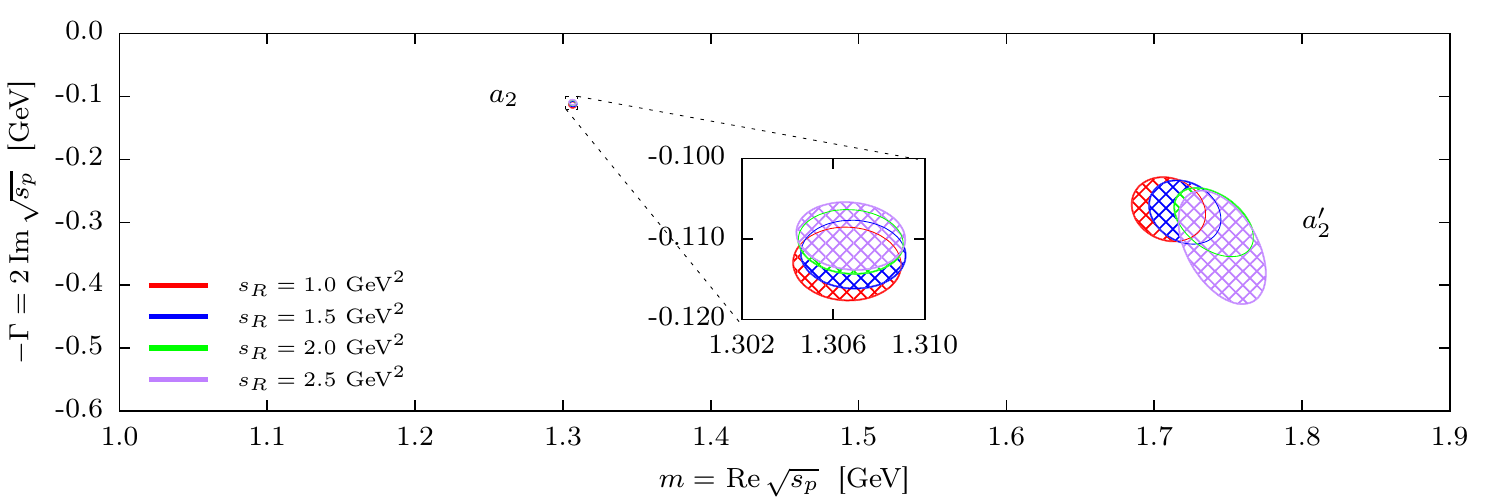}
\caption{Location of second-sheet pole positions  with two CDD poles, $n_p=6$, and with $s_R$ varied from $1.0$ GeV$^2$ to $2.5$ GeV$^2$. 
Poles are shown with $2\sigma$ ($95.5\%$) confidence level contours from uncertainties 
computed using  $10^5$ bootstrap fits. 
}
\label{fig:polepositions}  
\end{figure*}

This analysis allows us to extract the $\eta\pi\to\eta\pi$ elastic amplitude in the $D$-wave.  
By construction, the amplitude has a zero at $s=c_3$. 
Figure~\ref{fig:etapi} shows the real and imaginary parts of $\hat{f}(s)$, with the $3\sigma$ error bands estimated by the bootstrap analysis. 
Resonance poles are extracted by analytically continuing the denominator of the
$\eta\pi$ elastic amplitude to the second Riemann sheet (II)
across the unitarity cut using 
$D_{\text{II}}(s) = D(s) + 2i\rho(s)N(s)$. 
By construction, no first-sheet poles are present.
We find three second-sheet poles in the energy range of 
$(m_{\pi}+m_{\eta})\le \sqrt{s} \le 3$~GeV, two of which can be identified as resonances,
as shown in Fig.~\ref{fig:polepositions} 
for $n_p=6$ and $s_R=\{1.0, 1.5, 2.0, 2.5\}$~GeV$^2$. 

The mass and width are defined as $m=\re \sqrt{s_p}$ 
and $\Gamma=-2\,\im\sqrt{s_p}$, respectively, where $s_p$ is the pole position in the $s$ plane. 
Two of the poles found 
can be identified as the $a_2(1320)$ and $a_2^\prime(1700)$ resonances
in the PDG~\cite{pdg}.
The lighter of the two corresponds to the $a_2(1320)$. 
For $s_R=1.5$~GeV$^2$, the pole has mass and width 
$m = (1307 \pm 1)$ MeV and $\Gamma = (112 \pm 1)$~MeV, respectively. The nominal value is the best fit pole position, and the uncertainty is the statistical deviation determined from the bootstrap. 
Values of $s_R$ between 1.0 and 2.5~GeV$^2$ lead to pole deviations of at most
$\Delta \,m =2$ MeV and $\Delta \,\Gamma =3$~MeV. 
The heavier pole corresponds to the excited $a_2^\prime(1700)$. 
For $s_R=1.5$~GeV$^2$, 
the resonance has mass and width $m = (1720 \pm 10)$~MeV and $\Gamma = (280 \pm 10)$~MeV, respectively. 
The maximal deviations for the different $s_R$ values are 
$\Delta \,m =40$~MeV and $\Delta \,\Gamma =60$~MeV.
The $a_2(1320)$ and $a_2^\prime(1700)$ poles (see Fig.~\ref{fig:polepositions}) 
are found to be stable under variations of $s_R$, which modulates the left hand cut.
As expected, there is a third pole that depends strongly on $s_R$ and reflects the singularity in $N(s)$ modeled as a pole.  
Its mass ranges from $1.4$ to $3.3$~GeV, 
and its width varies between $1.3$ and $1.8$~GeV  
as $s_R$ changes from $1$~GeV$^2$ to $2.5$~GeV$^2$. 
In the limit $g\to 0$, this pole moves to $-s_R$ as expected, 
while the other two migrate to the real axis above threshold~\cite{animation}.

Changing the number of expansion terms between $n_p=3$ and $n_p=7$
does not in any significant way affect the $a_2(1320)$ or $a_2^\prime(1700)$ pole positions.
The maximal deviations are $\Delta\,m(a_2)=5$~MeV, $\Delta\,\Gamma(a_2)=7$~MeV 
and $\Delta\,m(a_2')=40$~MeV, $\Delta\,\Gamma(a_2')=30$~MeV 
between three and seven terms in the $n(s)$ expansion.

\begin{figure*}
\centering
\subfigure[Coupled CDD parameterization]{
\includegraphics[width=.48\textwidth]{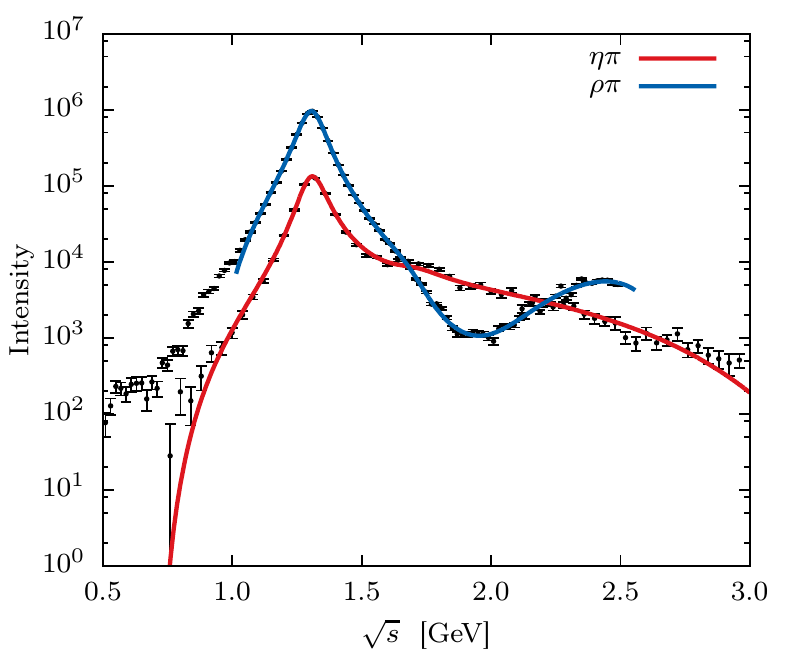}}
\subfigure[$K$-matrix]{
\includegraphics[width=.48\textwidth]{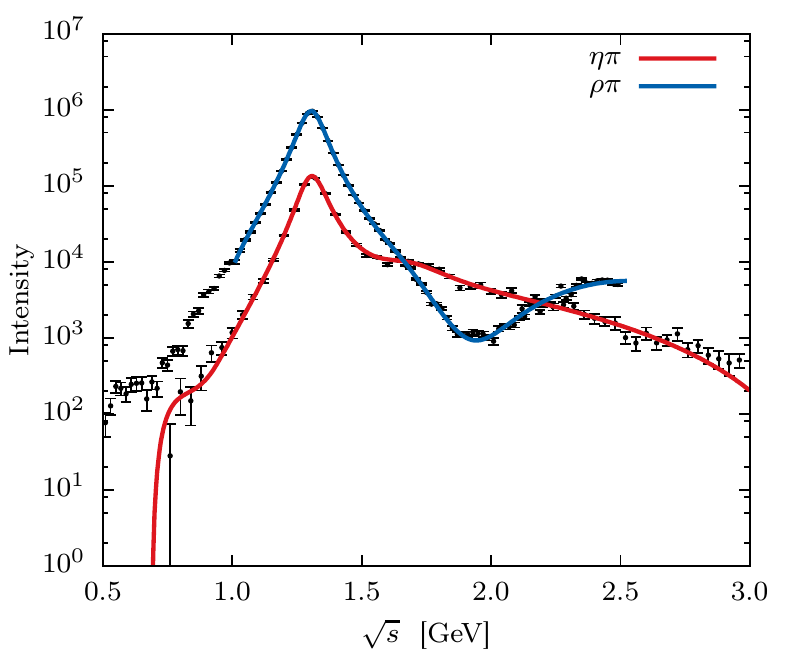}}
\caption{Coupled-channel $D$-wave fit, (a) 
using a model based on CDD poles, (b) using the standard $K$-matrix parameterization. 
Both parameterizations give pole positions consistent with the single-channel analysis. 
The $\eta\pi$ data is taken from Ref.~\cite{Adolph:2014rpp} 
and the $\rho\pi$ data from Ref.~\cite{Adolph:2015tqa}.}
\label{fig:ccfit}  
\end{figure*}

In order to demonstrate that coupled-channel effects do not influence the pole positions, we consider an extension of the model to include a second channel also measured by COMPASS, $\rho\pi$ \cite{Adolph:2015tqa}, 
and simultaneously fit the 
$\eta \pi$~\cite{Adolph:2014rpp} and the $\rho \pi$~\cite{Adolph:2015tqa} final states.
The branching ratio of the $a_2(1320)$ is saturated at the level of $\sim$85\% by the $\eta\pi$ and $3\pi$ channels~\cite{pdg}, with the 
 $\rho\pi\,$ $S$-wave having the dominant contribution.
 For simplicity we consider the $\rho$ to be a stable particle with mass 775 MeV, 
the finite width of the $\rho$ being relevant only for $\sqrt{s} < 1$~GeV. 
The amplitude is then $\hat{a}_{j}(s)=\sum_{k}\left[D(s)\right]^{-1}_{jk}(s)\: n_{k}(s)$.
The denominator is now a $2\times 2$ matrix, whose diagonal elements are of the form given by 
Eq.~\eqref{eq:Dsol}, with the appropriate phase space for each  channel. 
The off-diagonal term is parameterized as a single real constant. 
The production elements $n_k(s)$ are as in Eq.~\eqref{eq:neqn}, with independent 
coefficients for each channel. 
We also performed a $K$ matrix coupled-channel fit and obtained very similar results that are shown in 
Figure~\ref{fig:ccfit}. 
 The coupled-channel effects produce a competition 
between the parameters in the numerators to fit the bump at 1.6 GeV in $\eta\pi$  and the dip at $1.8$ GeV in $\rho\pi$ at the same time. 
The $\rho\pi$ data prefers not to have any excited $a_2'(1700)$, 
which conversely is evident in the $\eta\pi$ data. 
Therefore, the uncertainty in the $a_2'(1700)$ pole position increases, 
as it is practically unconstrained by the $\rho\pi$ data. 
Note, however, that in Ref.~\cite{Adolph:2015tqa} the dip at $\sqrt{s}\sim 1.8$ GeV in the $\rho\pi$ data is $t$-dependent, while we use the $t$-integrated  intensity, so it may be expected that the effects of the $a_2'$ are suppressed in our combined fit.

We find the following deviations in the pole positions relative 
to the single-channel fit: $\Delta m(a_2) = 2$~MeV, 
$\Delta \Gamma(a_2) = 3$~MeV, $\Delta m(a_2') = 20$~MeV 
and $\Delta \Gamma(a_2') = 10$~MeV. 
These deviations are rather small and we quote them within our systematic uncertainties.

%

\section{Summary and Outlook}
We describe the $2^{++}$ wave of $\pi p\to\eta\pi p$ reaction
in a single-channel analysis emphasizing unitarity and analyticity of the amplitude.
These fundamental $S$-matrix principles significantly constrain the possible form of the amplitude making the analysis more 
stable 
than standard ones that use 
sums of Breit-Wigner resonances with phenomenological background terms.

The robustness of the model allows us to reliably reproduce the data, 
and to extract pole positions by analytical continuation to the complex $s$-plane.  
We use the single-energy partial waves in Ref.~\cite{Adolph:2014rpp}
to extract the pole positions. 
We find two poles that can be identified as the $a_2(1320)$ and 
the $a_2^\prime(1700)$ resonances, with pole parameters
\begin{align*}
m({a_2}) &= (1307 \pm 1 \pm 6) \text{ MeV}, 
& m({a'_2}) &= (1720 \pm 10 \pm 60) \text{ MeV}, \\
\Gamma({a_2}) &= \phantom{0}(112 \pm 1 \pm 8)\text{ MeV}, 
& \Gamma({a_2^\prime}) &= \phantom{0}(280 \pm 10 \pm 70)\text{ MeV},
\end{align*}
where the first uncertainty is statistical (from the bootstrap analysis)
and the second one systematic. 
The systematic uncertainty is obtained 
adding in quadrature the different systematic effects, \ie
the dependence on
the number of terms in the expansion of the numerator function $n(s)$,
on $s_R$, on $t_\text{eff}$ (negligible), and on the coupled-channel effects. 
The $a_2$ results are consistent with the previous $a_2(1320)$ results found in Ref.~\cite{Adolph:2014rpp}. We note that a new mass-dependent COMPASS analysis of the $3\pi$ final state using Breit-Wigner forms in 14 waves is in progress. 

The third pole found tends to $-s_R$ in the limit of vanishing coupling, indicating that this pole arises from the treatment of the  exchange forces, 
and not from the CDD poles that account for the resonances. 

  In the future this analysis will be extended to also include the 
   $\eta^{\prime}\pi$ channel~\cite{jpacinpreparation}, where a large exotic $P$-wave is observed \cite{Adolph:2014rpp}.

Additional material is available online through an interactive website~\cite{Mathieu:2016mcy,JPACweb}.

\section*{Acknowledgments}
We acknowledge the Lilly Endowment, Inc., 
 through its support for the Indiana University Pervasive Technology Institute, 
and the Indiana METACyt Initiative. 

\bibliographystyle{elsarticle-num}
\bibliography{quattro}
\end{document}